\documentclass[article]{aa}
\usepackage{natbib}         
\usepackage{graphicx,epstopdf}
\usepackage{longtable}
\usepackage[switch]{lineno} 
\usepackage{xcolor}
\usepackage[flushleft]{threeparttable}
\usepackage{siunitx} 
\bibpunct{(}{)}{;}{a}{}{,} 

\newcommand{\ind}[1]{_{\mathrm{#1}}}

\newcommand\numax{\nu\ind{max}}

\newcommand\Teff{T\ind{eff}}

\newcommand{\mean}[1]{\overline{#1}}

\begin{document}

\author{%
Charlotte Gehan\inst{1},
Patrick Gaulme\inst{1},
Jie Yu\inst{1}}

\institute{Max-Planck-Institut für Sonnensystemforschung, Justus-von-Liebig-Weg 3, 37077 Göttingen, Germany ; \texttt{gehan@mps.mpg.de}}

\abstract{According to dynamo theory, stars with convective envelopes efficiently generate surface magnetic fields, which manifest as magnetic activity in the form of starspots, faculae, flares, when their rotation period is shorter than their convective turnover time. Most red giants, having undergone significant spin down while expanding, have slow rotation and no spots. However, based on a sample of about 4500 red giants observed by the NASA \textit{Kepler} mission, a previous study showed that about 8\,\% display spots, including about 15\,\% that belong to close binary systems. Here, we shed light on a puzzling fact: for rotation periods less than 80 days, a red giant that belongs to a close binary system displays a photometric modulation about an order of magnitude larger than that of a single red giant with similar rotational period and physical properties. 
We investigate whether binarity leads to larger magnetic fields when tides lock systems, or if a different spot distribution on single versus close binary stars can explain this fact. For this, we measure the chromospheric emission in the Ca\textsc{ii} H \& K lines of 3130 of the 4465 stars studied in a previous work thanks to the LAMOST survey. We show that red giants in a close-binary configuration with spin-orbit resonance display significantly larger chromospheric emission than single stars, suggesting that tidal locking leads to larger magnetic fields at a fixed rotational period. Beyond bringing interesting new observables to study the evolution of binary systems, this result could be used to distinguish single versus binary red giants in automatic pipelines based on machine learning.}


\title{Surface magnetism of rapidly rotating red giants:\\single versus close binary stars \footnote{Table 1 is only available in electronic form at the CDS via anonymous ftp to cdsarc.u-strasbg.fr (130.79.128.5) or via \url{https://cdsarc.cds.unistra.fr/cgi-bin/qcat?J/A+A/}}}
\authorrunning{C. Gehan, P. Gaulme, J. Yu}
\titlerunning{Magnetic activity of red giants: single versus binary stars}
\maketitle

\section{Introduction}\label{introduction}
During the main sequence, stars with masses less than $1.3M_\odot$ have an external convective envelope that produces a complex dynamo mechanism, whose efficiency depends on the interaction between differential rotation and subphotospheric convection \citep{Skumanich_1972}. According to dynamo theory, surface magnetic fields are efficiently generated when the stellar rotation period is shorter than the convective turnover time \citep[e.g.,][]{Charbonneau_2014}. Magnetic fields play an important role in stellar evolution. They are believed to regulate the stellar rotation from early to late stages of low-mass stars \citep{Vidotto_2014}. Additionally, magnetic fields create a strong coupling between the star and the stellar wind, which leads to loss of angular momentum \citep[e.g.,][]{Gallet_2013} during the main sequence, and an important mass loss during the red giant phase \citep{Harper_2018}.

Magnetic fields generate various phenomena when emerging from the outer convective envelope, which are grouped under the name of stellar magnetic activity \citep{Hall_2008}. Studying stellar activity thus provides information about the physical dynamo process. It is also an important observable for exoplanetary science because it impacts habitability, and hampers the detection of planetary signals \citep{Borgniet_2015}. 

Magnetic activity is a function of stellar evolution, with stars spinning down and becoming less active as they evolve \citep[e.g.,][]{Wilson_1964, Skumanich_1972}. However, based on a sample of 4465 red giants observed by the NASA \textit{Kepler} mission, \cite{Gaulme_2020} showed that about 8\,\% display a photometric rotational modulation caused by stellar spots. Among these 8\,\%, about 85\,\% are single stars, whereas 15\,\% belong to close binary systems, typically with orbits shorter than 150 days. Red giants in close binaries are expected to display a magnetic activity: as they evolve along the red giant branch (RGB), the components can reach a tidal equilibrium where stars are synchronized and orbits circularized \citep[e.g.][]{Verbunt_1995, Beck_2018}. Red giant stars are forced to spin faster than what they would if isolated, which results in surface magnetic fields, hence spots, sometimes flares, and a larger photometric modulation. Such a type of close binaries including an active evolved star is usually classified as RS CVn stars, from the prototype member RS Canum Venaticorum \citep[e.g.][]{Hall_1976, Strassmeier_1988}. This phenomenon is associated with a suppression of the solar-like oscillations, which can be partial or total \citep{Gaulme_2014, Gaulme_2016, Beck_2018, Benbakoura_2021}.

However, the picture is probably more complex than it seems. Indeed, \cite{Gaulme_2020} noticed that at a given rotation period, below 80 days, a red giant that belongs to a close binary system displays a photometric modulation about one order of magnitude larger than that of a single red giant with similar rotation period and physical properties (mass, radius). This observation could be explained in two different manners: either tidal locking somehow leads to larger magnetic fields, or the spot distribution differs between binary and single red giants, leading to a different photometric variability. For example, a single large spot can lead to a larger photometric amplitude than a series of smaller spots at all longitudes. It has been shown that stars in RS CVn systems can exhibit special spot distributions, with the presence of active longitudes synchronised with the orbital period in the direction of the line of centres in the binary \citep[e.g][]{Berdyugina_1998, Kajatkari_2014}. This can in turn result in larger photometric amplitudes than observed for single stars exhibiting different spot distributions. In such case, the larger photometric amplitudes measured for red giants in close binaries are not necessarily due to larger magnetic fields. Unfortunately, photometry alone cannot solve this question. 

In this work, we aim at figuring out whether red giants in close binary systems with a spin-orbit resonance have a larger magnetic field than single red giant with similar physical and rotational properties. For this, we measure the chromospheric emission in the Ca\textsc{ii} H \& K lines $S\ind{CaII}$ (a.k.a. $S$-index), which is a proxy of the strength of surface magnetic fields, from the spectra of the Large Sky Area Multi-Object Fibre Spectroscopic Telescope survey \citep[LAMOST,][]{Cui_2012}. We work with the sample of 4465 red giants studied by \citet{Gaulme_2020}.

In Sect. \ref{Sect_Sindex}, we describe the method we employed to measure chromospheric emission indices from LAMOST spectra. In Sect. \ref{Sect_SCa_Sph}, we study the correlation between $S\ind{CaII}$ and the  photometric index $S\ind{ph}$, which is proportional to the amplitude of the photometric variability. Then, we present $S\ind{CaII}$ for 3130 red giants measured from LAMOST spectra (Sect. \ref{Sect_Sindex_RGs}). In Sect. \ref{Sect_Sindex_binary_single}, we compare the index $S\ind{CaII}$ between single and binary stars. Finally, we investigate the impact of mass gain on $S\ind{CaII}$ for stars exhibiting signatures of mass transfer and stellar merging (Sect. \ref{Sect_mass-transfer}), before concluding (Sect. \ref{Sect_conclusion}).


\section{Method and Data}
\label{Sect_Sindex}

\subsection{Index of chromospheric emission}
The chromospheric activity encompasses a wide range of phenomena that produce non-thermal excess emission with respect to a radiative equilibrium atmosphere \citep{Hall_2008}, and is a proxy of the strength of surface magnetic fields \citep{Babcock_1961, Petit_2008, Auriere_2015, Brown_2022}. The chromospheric activity of cool stars is usually reported in the form of the $S$-index, which is a measure of the emission-line cores of the Ca\textsc{ii} H and K lines that are centred at 3968.470 \AA \, and 3933.664 \AA, respectively. A famous long-term survey of stellar surface magnetism is the The Mount Wilson project \citep{Baliunas_1995, Duncan_1991}, which reported $S$-indices for a large sample of cool stars from 1966 to the early 2000s.

The $S$-index $S\ind{CaII}$ is the flux ratio of the Ca\textsc{ii} K \& H spectral lines to two nearby bandpasses \citep{Wilson_1978,Duncan_1991, Karoff_2016, Zhang_2020, Gomes_2021}:
\begin{equation}
S\ind{CaII} = \frac{F\ind{H} + F\ind{K}}{F\ind{B} + F\ind{R}},
\end{equation}
where $F\ind{H}$ and $F\ind{K}$ correspond to the integrated flux in the Ca\textsc{ii} H and K lines, and $F\ind{B}$ and $F\ind{R}$ to the integrated flux in the blue and red regions of the pseudo-continuum centred at 3901.070 \AA \, and 4001.070 \AA, respectively.

We note that \cite{Gomes_2021} also work with the modified index of chromospheric emission $R^{'}\ind{HK}$, which essentially gives the fraction of stellar bolometric luminosity radiated in the form of chromospheric H and K emission \citep{Hall_2008}, by subtracting the flux in the H and K lines wings, which is mainly from photospheric origin. This parameter was introduced to compare chromospheric activity between stars with different effective temperatures $\Teff$. However, since we focus on low-RGB and RC stars, whose temperatures range from 4700 to 5200 K, we can safely use the $S\ind{CaII}$ to compare the chromospheric activity within our sample.

\subsection{From a spectrum to S-index}
The first step consists of determining the line-of-sight velocity (a.k.a. radial velocity) of the considered target before measuring the chromospheric emission. For this, we select a broad region of the spectrum (6530 -- 6590 \AA) surrounding the H$\ind{\alpha}$ line, which is located at $\lambda\ind{H\ind{\alpha},0}$ = 6562.801 \AA\ in absence of Doppler effect. We then find the actual center of the $H\ind{\alpha}$ line as the wavelength associated to the flux minimum $\lambda\ind{H\ind{\alpha}}$ to retrieve the stellar radial velocity $v\ind{r} = c \, \delta\lambda\ind{H\ind{\alpha}}/\lambda\ind{H\ind{\alpha},0}$, where $c$ is the speed of light in the vacuum and $\delta\lambda\ind{H\ind{\alpha}} = \lambda\ind{H\ind{\alpha}} - \lambda\ind{H\ind{\alpha},0}$. The Doppler-shifted wavelength of the Ca\textsc{ii} H and K lines is given by:
\begin{equation}
\lambda\ind{CaII,X} = \lambda\ind{CaII,X\ind{0}} \left ( 1 + \frac{v\ind{r}}{c} \right ),
\end{equation}
where X = {H,K}.

The flux in the CaII H and K lines is then integrated using a triangular bandpass, while the flux in the blue and red reference regions is integrated with a square bandpass having a width of 20 \AA. We use the same approach as described in \cite{Gomes_2021}. The full width at half maximum (FWHM) of the triangular bandpass depends on the resolution of the spectrograph, since it affects the precision one can get on the radial velocity measurement. Hence, we use a FWHM of 1.09 \AA \, for HARPS data as in \cite{Gomes_2021}, while we use a FWHM twice larger (2.18 \AA) for LAMOST data, to ensure that the centre of the CaII H and K lines is included in the integrated flux.

Uncertainties are computed from $\mean{SNR}$, which is the mean value of the SNR over a wavelength range encompassing the CaII H and K lines as well as the blue and red regions of the pseudo-continuum between 3881.07\,\AA \, and 4021.07\,\AA \, such as \citep{Karoff_2016}
\begin{equation}\label{eqt-sigma}
\log\sigma\ind{S\ind{CaII}} = -\log{\mean{SNR}} - 0.5.
\end{equation}
The SNR is computed as
\begin{equation}\label{eqt-SNR}
\mathrm{SNR} = F  \sqrt{I\ind{v}},
\end{equation}
where $F$ is the flux and $I\ind{v}$ is the inverse variance of the flux, provided in the LAMOST spectra files. Note that in case of pure photon noise, Eq. \ref{eqt-SNR} turns into $\mathrm{SNR} = \sqrt{F}$.

Because the Ca\textsc{ii} H and K lines are located in the blue, where the SNR is lower than at higher wavelengths, the background light correction can result in a nonphysical negative flux. We follow the recommendations of \citet{Zhang_2020} and chose to keep the spectra for which less than 1\% of the spectrum has negative values in the K and H bandwidths, as well as $\mean{SNR} > 3$ as in \cite{Gomes_2021}. We tolerate 1\,\% of negative flux because low SNR spectra may have stochastic spikes with negative values but remain clearly positive in average. 

When several spectra are available for a given star, we compute the S-index as the median of all the S-indices measured for each individual spectrum. Our method delivers S-index measurements in one to two seconds of computational time on a regular laptop computer.

\subsection{Validation on HARPS spectra}\label{HARPS}

\begin{figure}
\centering
\includegraphics[width=9.1cm]{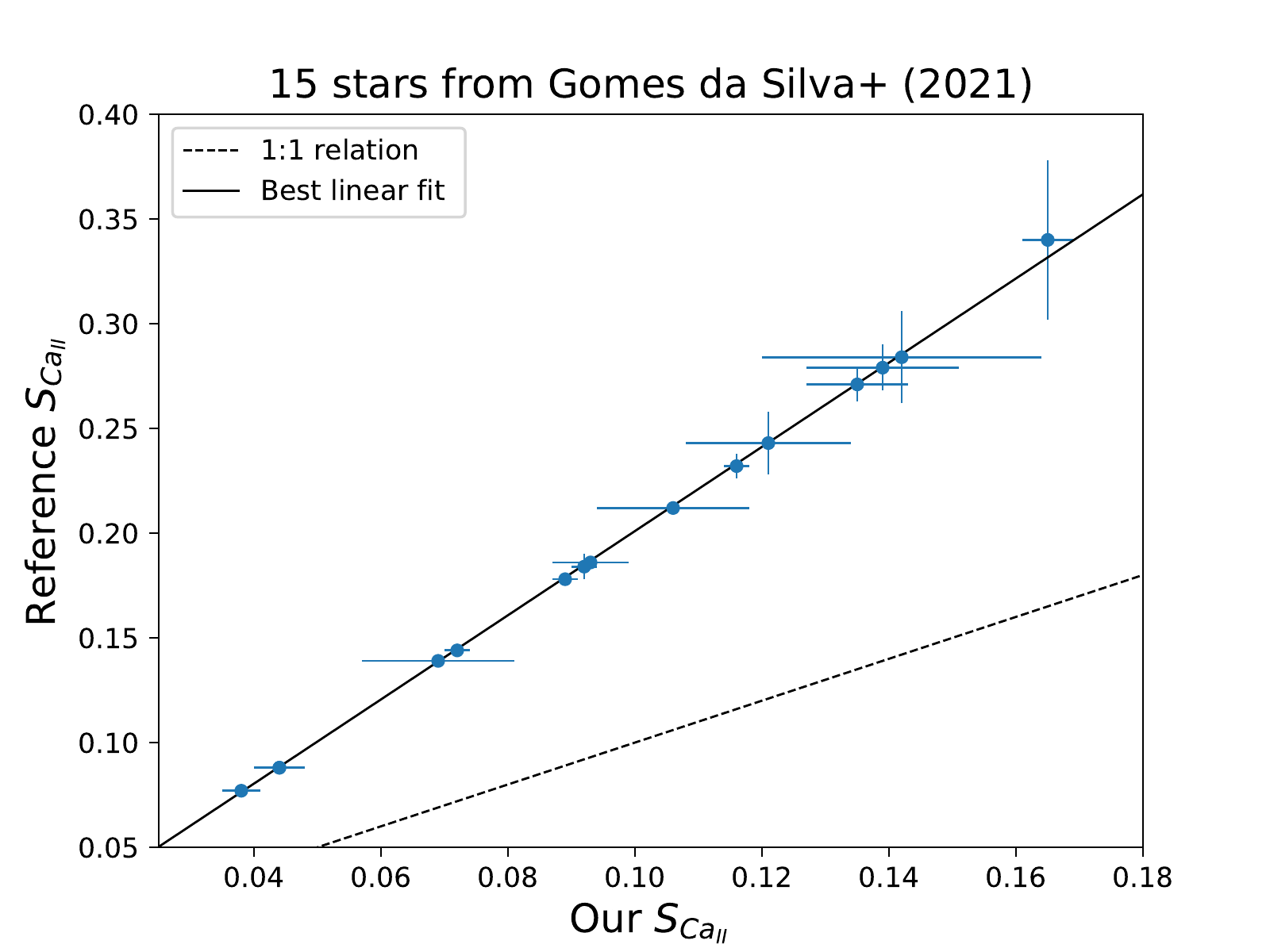}
\caption{Comparison between our S-index measurements and those from \cite{Gomes_2021} for 15 FGK stars using HARPS data. The dotted lines represent a 1:1 comparison while the continuous line represents the best linear fit with a coefficient $a = 2.01 \pm 0.01$.}
\label{fig-HARPS}
\end{figure}

We first tested our routine on FGK stars analysed by \cite{Gomes_2021} using spectroscopic data from the High Accuracy Radial velocity Planet Searcher (HARPS). We made sure to use the same exact spectra as \cite{Gomes_2021} to derive their S-index measurements. Therefore, we limited our analysis to 15 stars, each having a number of HARPS spectra comprised between 1 and 13. We obtain a perfectly linear relation between our measurements and those of \cite{Gomes_2021} (Fig. \ref{fig-HARPS}). We performed a linear fit and found that our measurements need to be multiplied by a factor of 2.01 $\pm$ 0.01 to match those of \cite{Gomes_2021}. This factor comes from the calibration of the S-index to the Mount Wilson scale that is usually performed. This factor is usually on the order of $\sim 1.8$ \citep{Karoff_2016}, which is globally consistent with what we obtain.

\begin{figure}
\centering
\includegraphics[width=9.1cm]{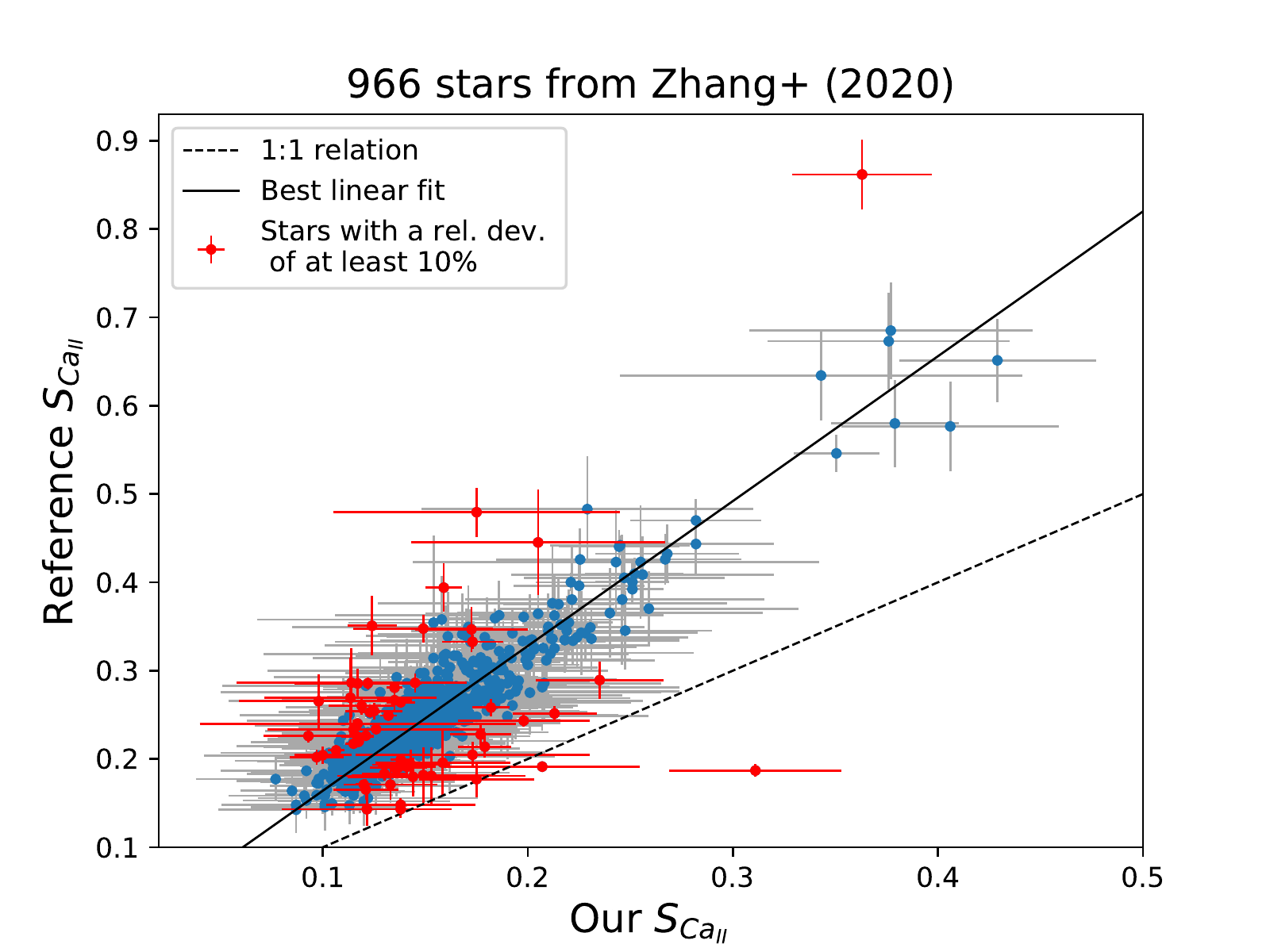}
\caption{Same as Fig. \ref{fig-HARPS}, but for 966 red giants from \cite{Zhang_2020} using LAMOST data. The best linear fit corresponds to a coefficient $a = 1.64 \pm 0.01$. Stars with a relative deviation below 10\% between S-index measurements are represented in blue, with error bars in grey for visualization purposes. Stars with a relative deviation of at least 10\% between S-index measurements are represented in red, as well as the associated error bars.}
\label{fig-LAMOST}
\end{figure}

\subsection{Validation on LAMOST spectra}\label{LAMOST}
We additionally applied our method to 1000 red giants analyzed by \cite{Zhang_2020} using spectroscopic data from the Large Sky Area Multi-Object Fibre Spectroscopic Telescope (LAMOST), which  has already obtained millions of stellar spectra for cool stars in the Milky Way, including many \textit{Kepler} targets \citep[e.g.,][]{Liu_2015, De_Cat_2015, Zong_2018}.  We first retrieved the right ascension $\alpha$ and declination $\delta$ of the stars of interest from the Mikulski Archive for Space Telescopes (MAST). We then performed a systematic search of LAMOST spectra associated to these celestial coordinates, within a 6-arcsec radius, among DR4 LAMOST data as \cite{Zhang_2020}.

We obtain a clear linear trend between our measurements and those of \cite{Zhang_2020} (Fig. \ref{fig-LAMOST}). We performed a linear fit and found that our measurements need to be multiplied by a factor of 1.64 $\pm$ 0.01 to match those of \cite{Zhang_2020}. As for HARPS data (Section \ref{HARPS}), this factor is globally consistent with the factor of $\sim 1.8$ usually used to calibrate the S-indices to the Mount Wilson scale.

We obtain consistent measurements, i.e with a relative deviation below 10\% compared to \cite{Zhang_2020}, in 94\% of the cases (blue dots with grey error bars in Fig. \ref{fig-LAMOST}). We note that among the 1000 red giants we selected for which \cite{Zhang_2020} measured a S-index, we end up with only 989 stars with at least one LAMOST spectrum, while we are using LAMOST DR4 as \cite{Zhang_2020}. This indicates that we obtain different KIC identifications compared to \cite{Zhang_2020} in some cases, which results in inconsistencies in the measured S-index. 
Additionally, we were able to measure a S-index for 966 stars out of 989, indicating that \cite{Zhang_2020} used different criteria than us to decide whether to discard a spectrum with too low SNR. This is also a source of inconsistencies between our measurements, since we do not necessarily derive the S-index of a given star relying on the same exact spectra selected by \cite{Zhang_2020}. These two aspects most probably explain the 58 stars with inconsistent S-index measurement. 

\subsection{Sample analyzed}

We consider the seventh LAMOST data release (DR7) to look for spectra among the 4465 \textit{Kepler} red giants with radii between 4 and $15\,R_\odot$ that were analyzed by \cite{Gaulme_2020}. In absence of stellar names in the LAMOST catalog, we crossmatched the \textit{Kepler} and LAMOST targets thanks to their coordinates. We identified 3370 stars that have at least one LAMOST spectrum within a 6-arcsec radius on the sky. Then, we discarded 240 stars because their spectra had either nonphysical negative fluxes or exhibited a low SNR near the calcium lines. We measured chromospheric-emission indices $S\ind{CaII}$ of 3130 red giants. Table \ref{table:1} reports the S-index and associated uncertainty we measure for the 3130 red giants from \cite{Gaulme_2020} analyzed in this study (see also Appendix \ref{App_median_S_index}). The whole table  accessible on the Centre de Données de Strasbourg (CDS) database. Measurements of the photometric index $S\ind{ph}$, oscillations frequency at maximum amplitude $\numax$, height of the Gaussian envelope employed to model the oscillation excess power $H\ind{max}$, as well as rotation period $P\ind{rot} $ we use in the following were obtained by \citet{Gaulme_2020}.

\begin{table}
\centering
\caption{Chromospheric emission index $S\ind{CaII}$ of the 3130 stars from the \citet{Gaulme_2020} sample that have exploitable LAMOST spectra.}
\begin{tabular}{c c c}
\hline
KIC & $S\ind{CaII}$ & Error\\
\hline
1027337 & 0.152 & 0.043\\
1160789 & 0.119 & 0.017\\
1161618 & 0.119 & 0.007\\
1162746 & 0.118 & 0.006\\
1163453 & 0.108 & 0.008\\
1163621 & 0.187 & 0.040\\
1294122 & 0.144 & 0.025\\
1429505 & 0.115 & 0.009\\
1433593 & 0.151 & 0.005\\
1433730 & 0.108 & 0.005\\
1433803 & 0.141 & 0.003\\
1435573 & 0.125 & 0.020\\
1569842 & 0.162 & 0.006\\
... & ... & ...\\
\hline
\end{tabular}
\tablefoot{ The whole table is accessible on the Centre de Donn\'ees de Strasbourg (CDS) database.}
\label{table:1}
\end{table} 


\section{Chromospheric versus photometric indices}
\label{Sect_SCa_Sph}

\begin{figure}
\includegraphics[width=8.5cm]{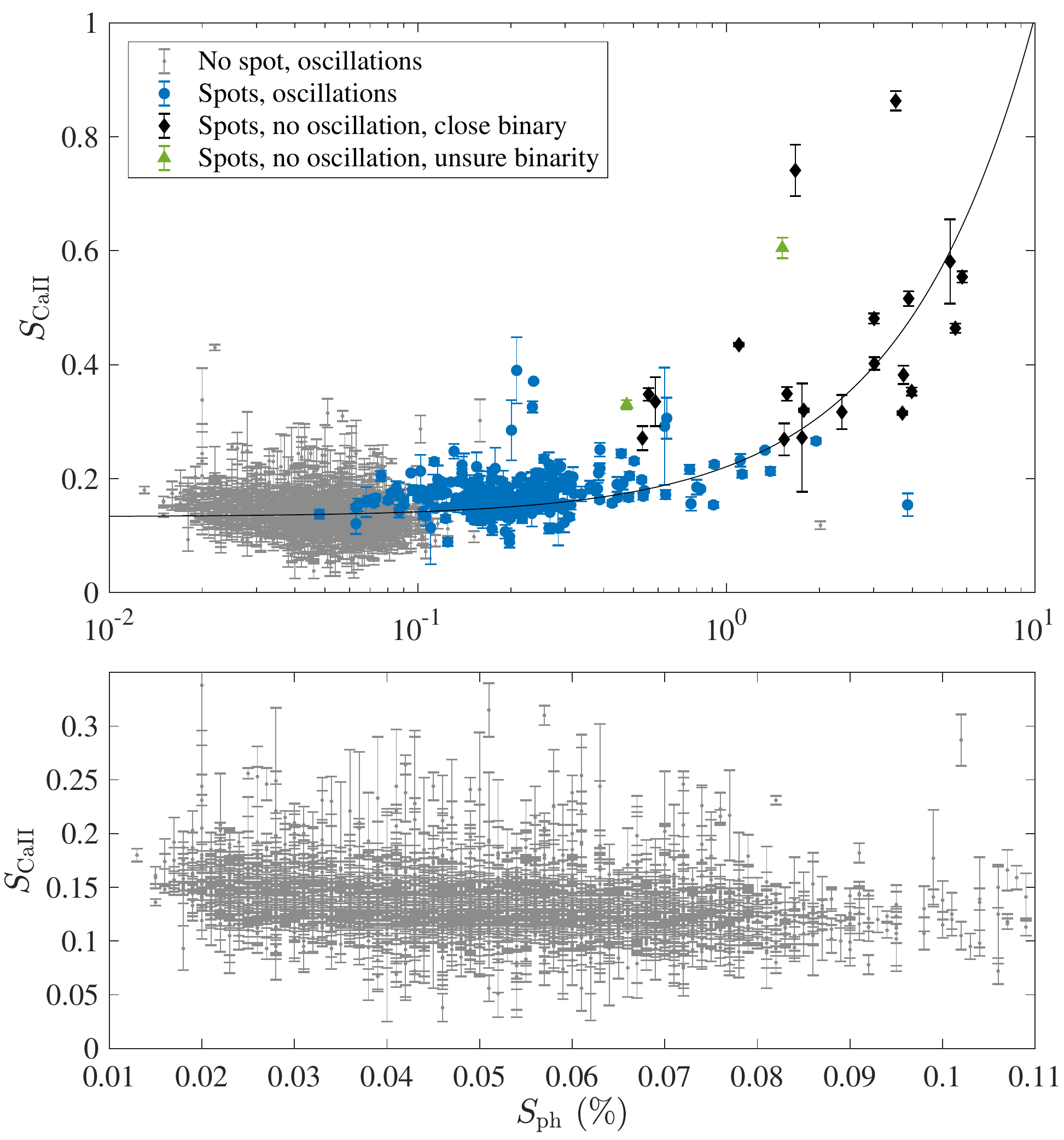}
\caption{Index of chromospheric emission $S\ind{CaII}$ as a function of photometric modulation $S\ind{ph}$ (in log scale). \textit{Upper panel:} Results for 3130 red giant stars observed by both LAMOST and \textit{Kepler}, from the \citet{Gaulme_2020} catalog. Gray symbols refer to regular inactive red giants, i.e. with no evidence of spot modulation, that display oscillations; the other colors refer to photometrically active giants, i.e. exhibiting spot modulation. Blue symbols refer to red giants with partially suppressed oscillations; black to non-oscillating red giants in close binary systems; green to non-oscillating red giants with ambiguous binary versus single status. The grey line represents the best linear fit given by Eq.~\ref{eqt-S-Sph}. Values of $S\ind{ph}$ are from \citet{Gaulme_2020}. \textit{Bottom panel:} Zoom on the upper panel for inactive red giants, with the $x$-axis in linear scale.}
\label{fig-Sph}
\end{figure}

\begin{figure*}
\includegraphics[width=9cm]{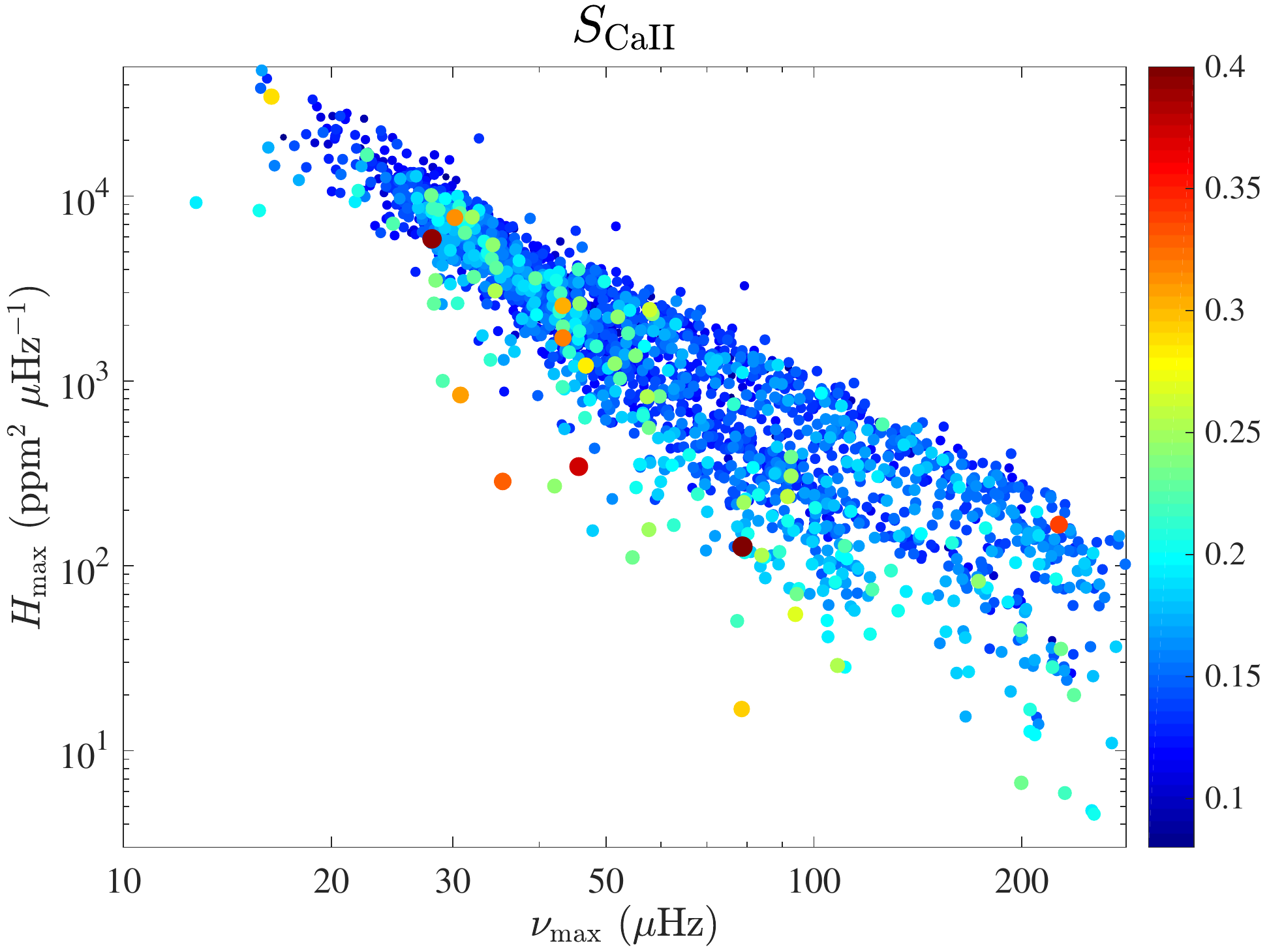}
\includegraphics[width=9cm]{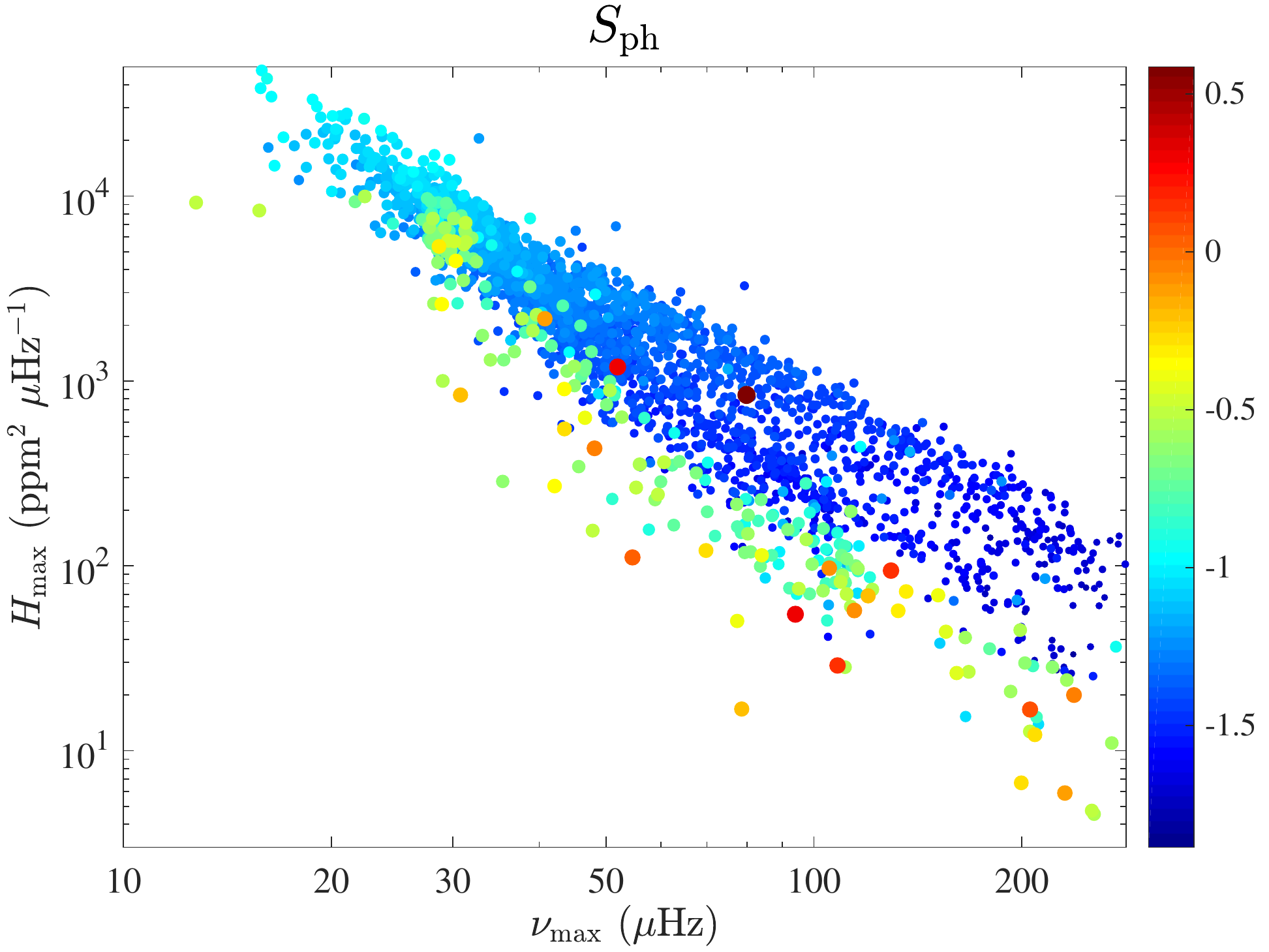}
\caption{\textit{Left:} Chromospheric emission index $S\ind{CaII}$ (colorscale and symbol size) as a function of oscillations frequency at maximum amplitude $\nu\ind{max}$ ($x$-axis) and height of the Gaussian envelope employed to model the oscillation excess power $H\ind{max}$ ($y$-axis) for the 3008 red giants observed both by \textit{Kepler} and LAMOST, for which $S\ind{ph}$ and $\nu\ind{max}$ are available. Colorscale was cut at 0.4 for readability purposes despite a few points reach $S\ind{CaII}\approx0.8$. \textit{Right:} corresponding figure for $S\ind{ph}$ as measured by \citet{Gaulme_2020}.}
\label{fig_S_CaII_Hmax_numax}
\end{figure*}

The photometric index $S\ind{ph}$ is defined as the standard deviation of the time series over five times the pseudo-period of the photometric modulation that is measured from the \textit{Kepler} lightcurves \citep{Mathur_2014}. When no photometric modulation is visible, $S\ind{ph}$ is the median of the lightcurve standard deviation over three-day intervals. Figure \ref{fig-Sph} displays $S\ind{CaII}$ as a function of $S\ind{ph}$ and reveals a clear global correlation between them, confirming that the photometric activity is proportional to the strength of surface magnetic fields. A linear fitting of $S\ind{CaII}$ as a function of $S\ind{ph}$ leads to:
\begin{equation}\label{eqt-S-Sph}
     S\ind{CaII} = 0.088_{-0.002}^{+0.002} S\ind{ph} + 0.133_{-0.001}^{+0.001}
\end{equation}
However, by looking closer, we observe a slight anticorrelation between $S\ind{CaII}$ and $S\ind{ph}$ for the sample of inactive stars, that is, those where \citet{Gaulme_2020} did not report any significant spot modulation after visually inspecting the light curves. We discuss that aspect in Sect. \ref{Sect_Sca_logg}. We also note that $S\ind{ph}$ has a much larger dynamical range of values, from about 0.02\,\% to about 8\,\%, whereas $S\ind{CaII}$ varies from about 0.05 to about 0.9. 

\begin{figure*}
\includegraphics[width=9cm]{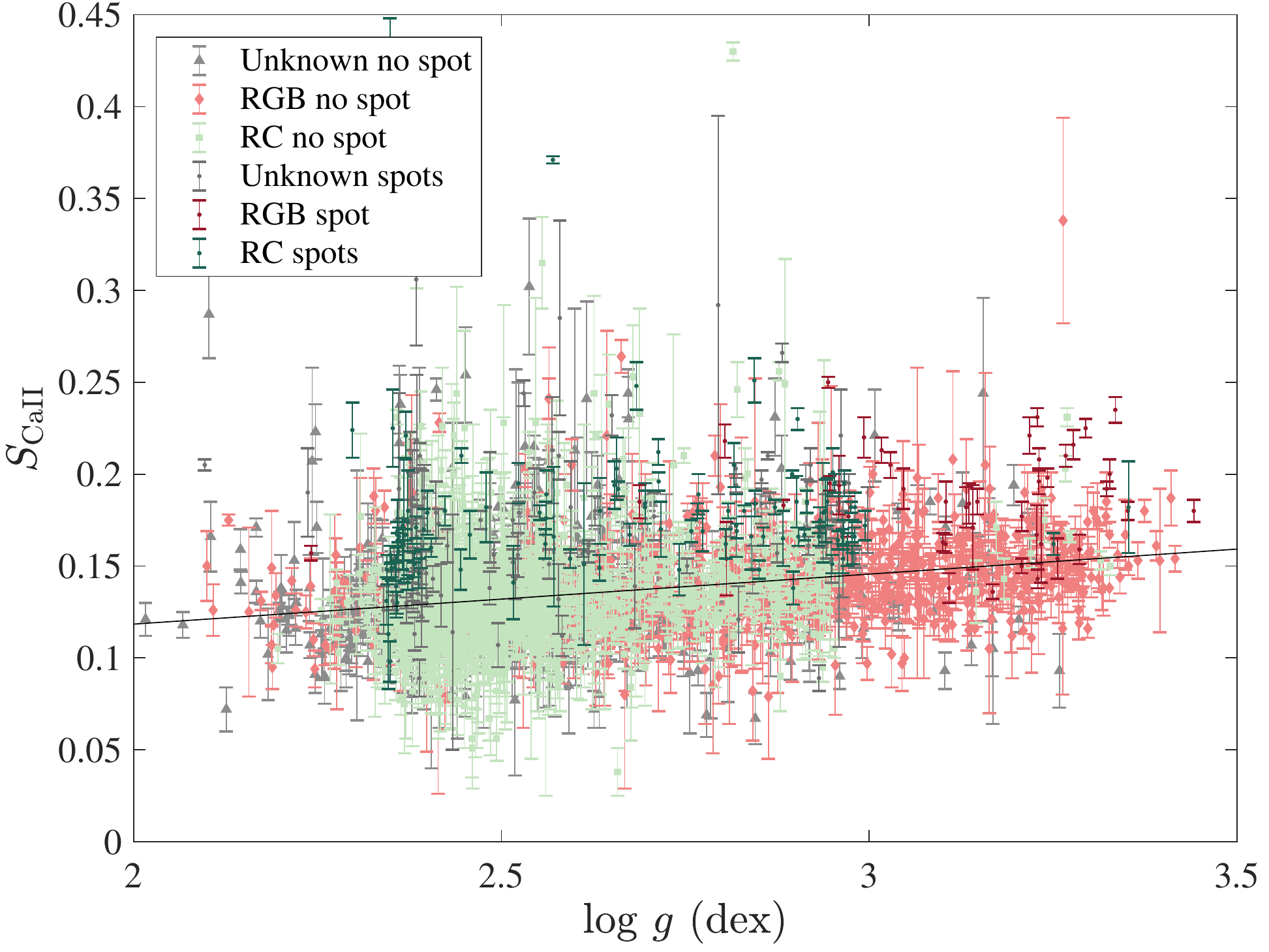}
\includegraphics[width=9cm]{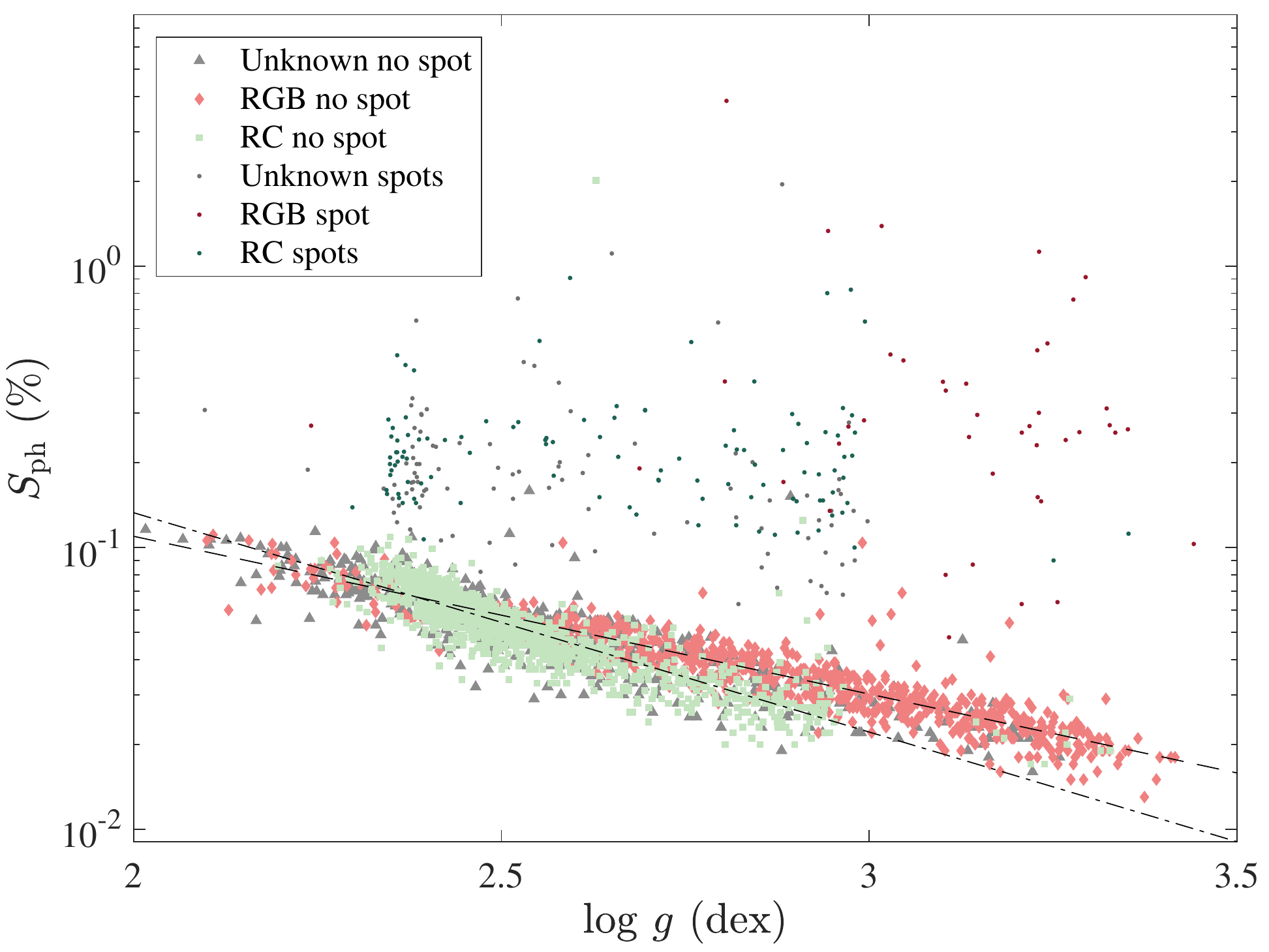}
\caption{\textit{Left:} Chromospheric emission index $S\ind{CaII}$ as a function of the surface gravity $\log g$. The color indicates the evolution stage; grey symbols refer to stars with an unknown evolution stage; red to RGB stars; green to red clump (RC) stars. Stars with spot modulation are represented by darker dots than stars with no spot modulation. The symbol changes as a function of evolution stage for stars with no spot modulation. The black line represents the best linear fit given by Eq.~\ref{eqt-S-log-g} computed for the red giants where no spot modulation was detected by \citet{Gaulme_2020}. \textit{Right:} Corresponding figure for $S\ind{ph}$ as measured by \citet{Gaulme_2020}. The two dashed black line represents the best linear fit of $S\ind{ph}$ for RGB (dashed) and RC (dot-dashed) stars given by Eq.~\ref{eqt-Sph-log-g}, computed for the red giants where no spot modulation was measured by \citet{Gaulme_2020}.}
\label{fig_logg_SCa}
\end{figure*}

\section{Chromospheric activity of red giants}
\label{Sect_Sindex_RGs}

\subsection{$S\ind{CaII}$ versus oscillation amplitude}\label{oscillations}
The magnetic field in stellar envelopes reduces the turbulent excitation of pressure waves by partially inhibiting convection. Since spots can also absorb acoustic energy, these two effects lead to a partial or total suppression of oscillations. \cite{Bonanno_2014} is the only work to directly confront oscillation amplitude and chromospheric emission, from a set of 19 solar-like pulsators on the main-sequence. They showed an anticorrelation between $S\ind{CaII}$ and the oscillation amplitude. The same was never checked for red giants. Our measurements of $S\ind{CaII}$ extend the trend observed by \cite{Bonanno_2014} to red giants, which is particularly visible for oscillators with frequencies at maximum amplitude $\numax > 55 \, \mu$Hz (Fig. \ref{fig_S_CaII_Hmax_numax}, left panel). This result is consistent with measurements of photometric modulation $S\ind{ph}$, where all of the active stars lay at the bottom of the oscillation amplitude diagram, indicating that red giants with weak oscillations display spots \citep[Fig. \ref{fig_S_CaII_Hmax_numax}, right panel, and][Fig. 7]{Gaulme_2020}. The trend we obtain for the S-index is not as clear as for $S\ind{ph}$, which likely results from the combination between relatively large errors on $S\ind{CaII}$ measurements, and the smaller range of values explored by $S\ind{CaII}$ with respect to $S\ind{ph}$.

\begin{figure*}[t]
\includegraphics[width=9.1cm]{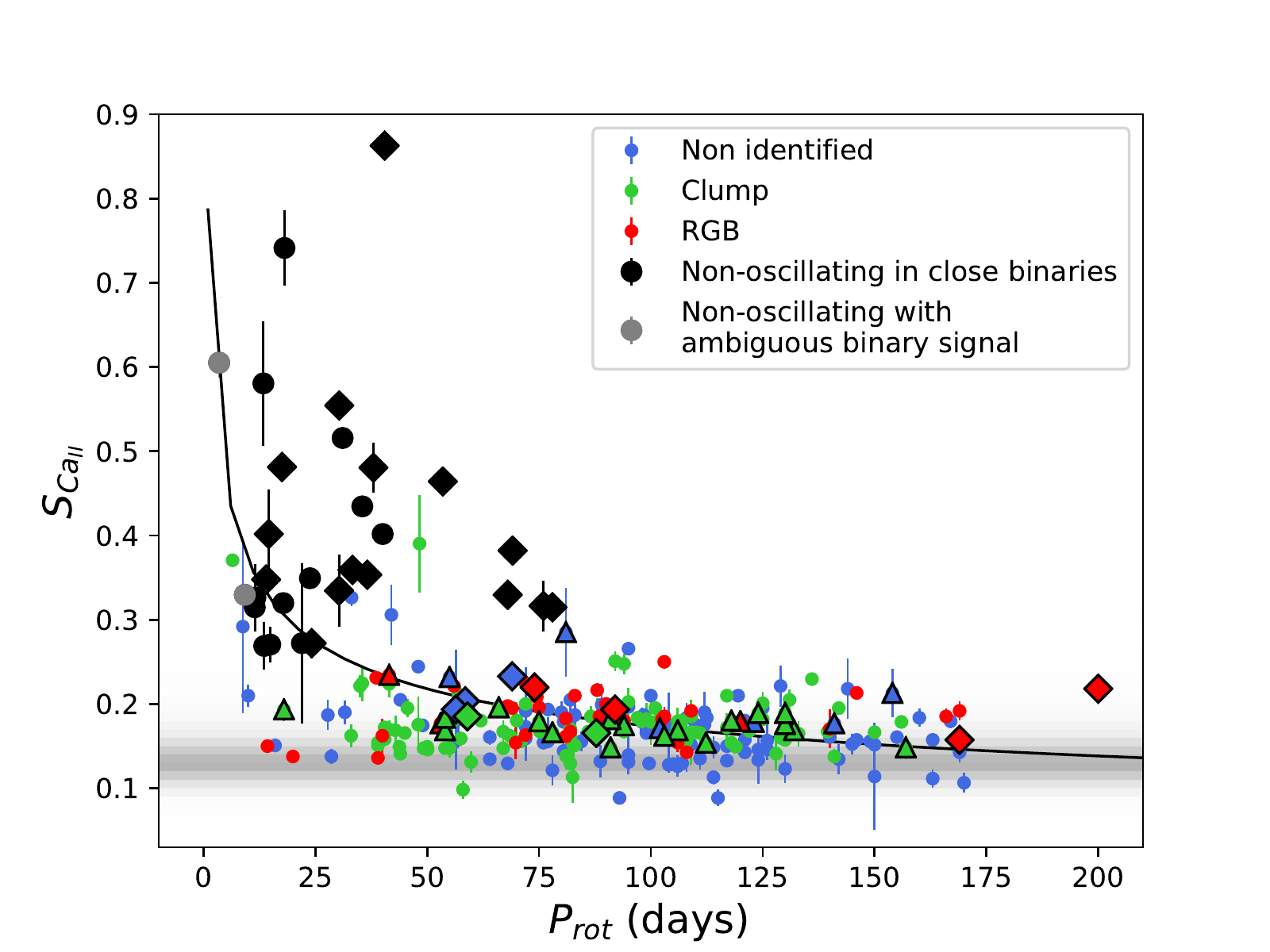}
\includegraphics[width=9.1cm]{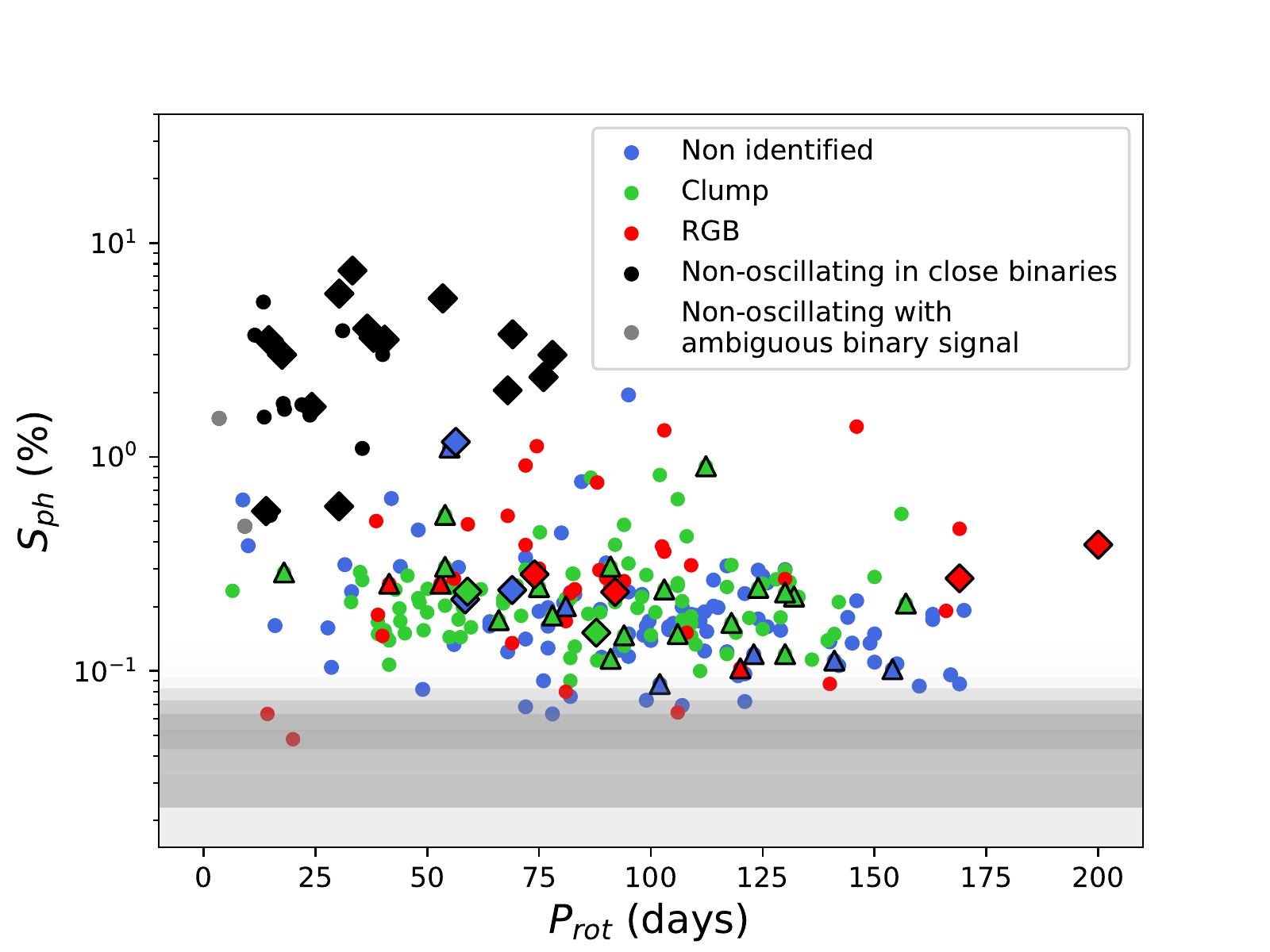}
\caption{\textit{Left:} Index of chromospheric emission $S\ind{CaII}$ as a function of rotational period $P\ind{rot}$. RGB and RC stars are represented in red and green, respectively, while red giants with an unidentified evolutionary stage are in blue. Black symbols correspond to non-oscillating red giants in close binaries. Gray dots indicate non-oscillating red giants with an ambiguous binary signal. Triangles and diamonds represent wide ($P\ind{orb}$ > 150 d) and close ($P\ind{orb} \leq$ 150 d) binaries identified by \textit{Gaia} DR3, respectively. The gray background indicates the distribution of $S\ind{CaII}$ for inactive stars, where the darker corresponds to the larger number of stars. The gray line represents the best fit of  $S\ind{CaII}$ as a function of $P\ind{rot}: S\ind{CaII} = 0.787(\pm0.070)\, P\ind{rot}^{-0.328 (\pm 0.023)}$. \textit{Right:} Corresponding figure for $S\ind{ph}$ as measured by \citet{Gaulme_2020}. It is a reproduction of Fig. 12 of \citet{Gaulme_2020} adapted to the sample for which we measured $S\ind{CaII}$.}
\label{fig-SCa-Prot}
\end{figure*}

\subsection{Chromospheric emission versus $\log g$}
\label{Sect_Sca_logg}
In Sect. \ref{Sect_SCa_Sph} (Fig. \ref{fig-Sph}), we noticed that $S\ind{CaII}$ and $S\ind{ph}$ are anticorrelated for stars that do not display spots. This fact is related to the way \citet{Gaulme_2020} computed $S\ind{ph}$: when no spot modulation is detected, $S\ind{ph}$ is the median standard deviation of the light curve over three day periods. It is thus much similar to the $F_8$ \textit{flicker} index introduced by \citet{Bastien_2013}, which measures the root mean square of the lightcurve on timescales shorter than 8 hours. \citet{Bastien_2013,Bastien_2016} demonstrated that $F_8$ is correlated with the amplitude of granulation and anticorrelated with the surface gravity. The latter is confirmed in Fig. \ref{fig_logg_SCa} (right panel), which shows a clear anticorrelation between $S\ind{ph}$ and asteroseismic $\log g$ (both from \citealt{Gaulme_2020}) for stars that are photometrically inactive, that is, where no spots are detected. We note a different slope for RGB and RC stars. A linear fitting of $S\ind{ph}$ as a function of $\log g$ for RGB and RC groups gives:
\begin{eqnarray}\label{eqt-Sph-log-g}
   S\ind{ph, RGB} &=& -0.560 \, (\pm0.017)\log g + 0.161 \,(\pm0.048)\\
   S\ind{ph, RC} &=& -0.779\, (\pm0.025)\log g + 0.682 \,(\pm0.063)
\end{eqnarray}
We note that the photometrically active red giants with oscillations do not follow this trend by having $S\ind{ph}$ values that stand out of the cohort of inactive stars.

In addition, it is well established that the oscillation amplitude is proportional to the granulation amplitude \citep[][]{Kallinger_2014}. Following the anticorrelation between the oscillation amplitude and the chromospheric index, it is normal to find that $S\ind{CaII}$ is proportional to the surface gravity (see Fig. \ref{fig_logg_SCa}, left panel). That being said, it is remarkable that despite the rather large noise of individual $S\ind{CaII}$ measurements, a clear trend is visible among the inactive stars. It means that even in absence of any noticeable surface magnetism -- no spot detected --, the chromospheric emission increases as a function of $\log g$. A linear fitting of $S\ind{CaII}$ as a function of $\log g$ performed on the stars with no spots leads to:
\begin{equation}\label{eqt-S-log-g}
     S\ind{CaII} = 0.027_{-0.003}^{+0.004} \log g + 0.064_{-0.010}^{+0.010}
\end{equation}
A red giant star with $\log g=2$ has in average an $S\ind{CaII}\approx0.12$, whereas a star with $\log g = 3.5$ has $S\ind{CaII}\approx0.16$.
A correlation between chromospheric emission and $\log g$ was already established by \citet[][their Fig. 10]{Gomes_2021} from a sample of stars with $\log g$ ranging from 2 to 5. \citet{Gomes_2021} used $R'\ind{HK}$ as an indicator of chromospheric activity. No correlation was directly visible among their RGs because of a rather small number of them. To the extent of our knowledge, this is the first time that a direct correlation between $S\ind{CaII}$ and $\log g$ is established for quiet red giants.

\begin{figure*}[t]
\includegraphics[width=9.1cm]{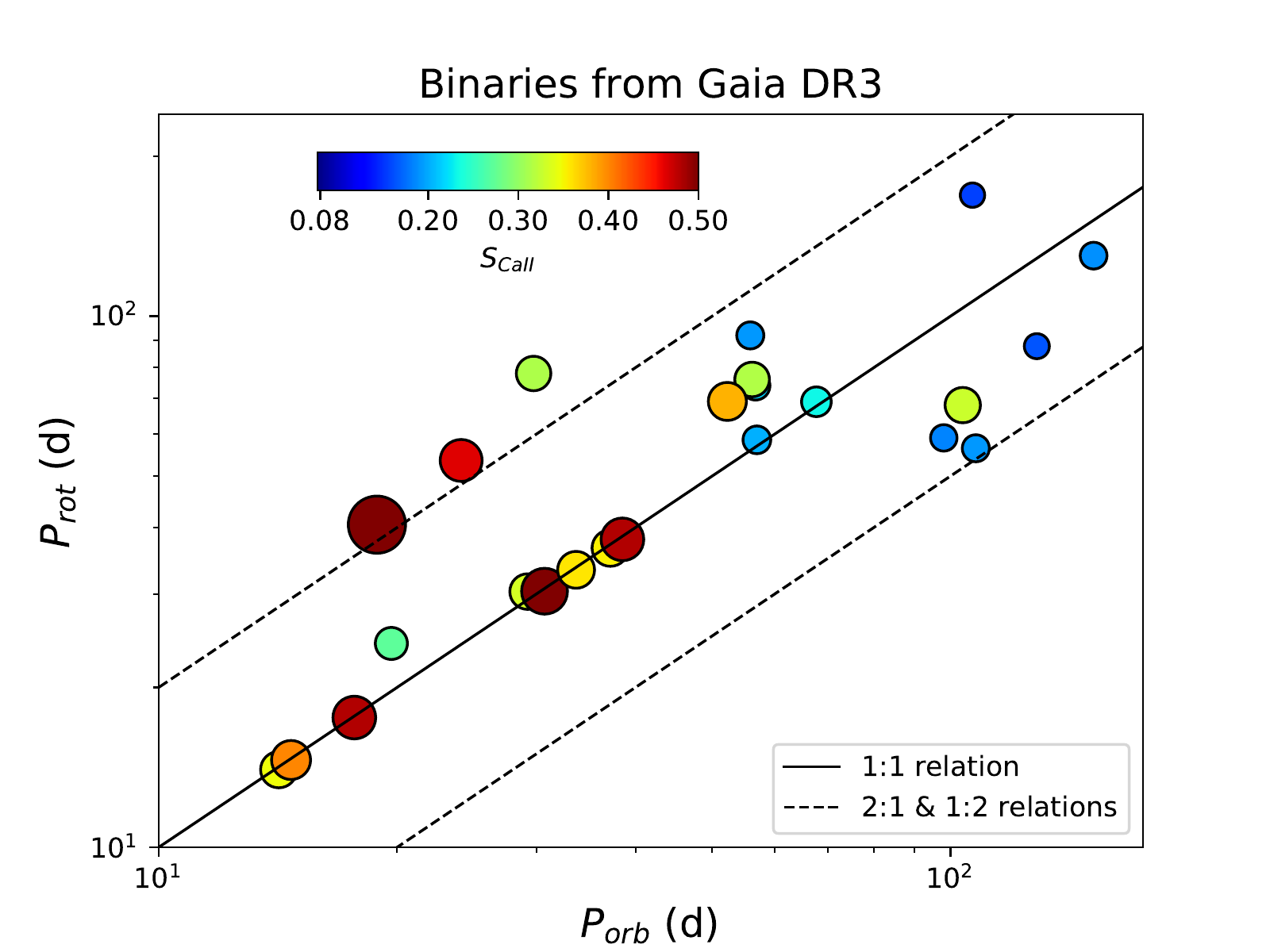}
\includegraphics[width=9.1cm]{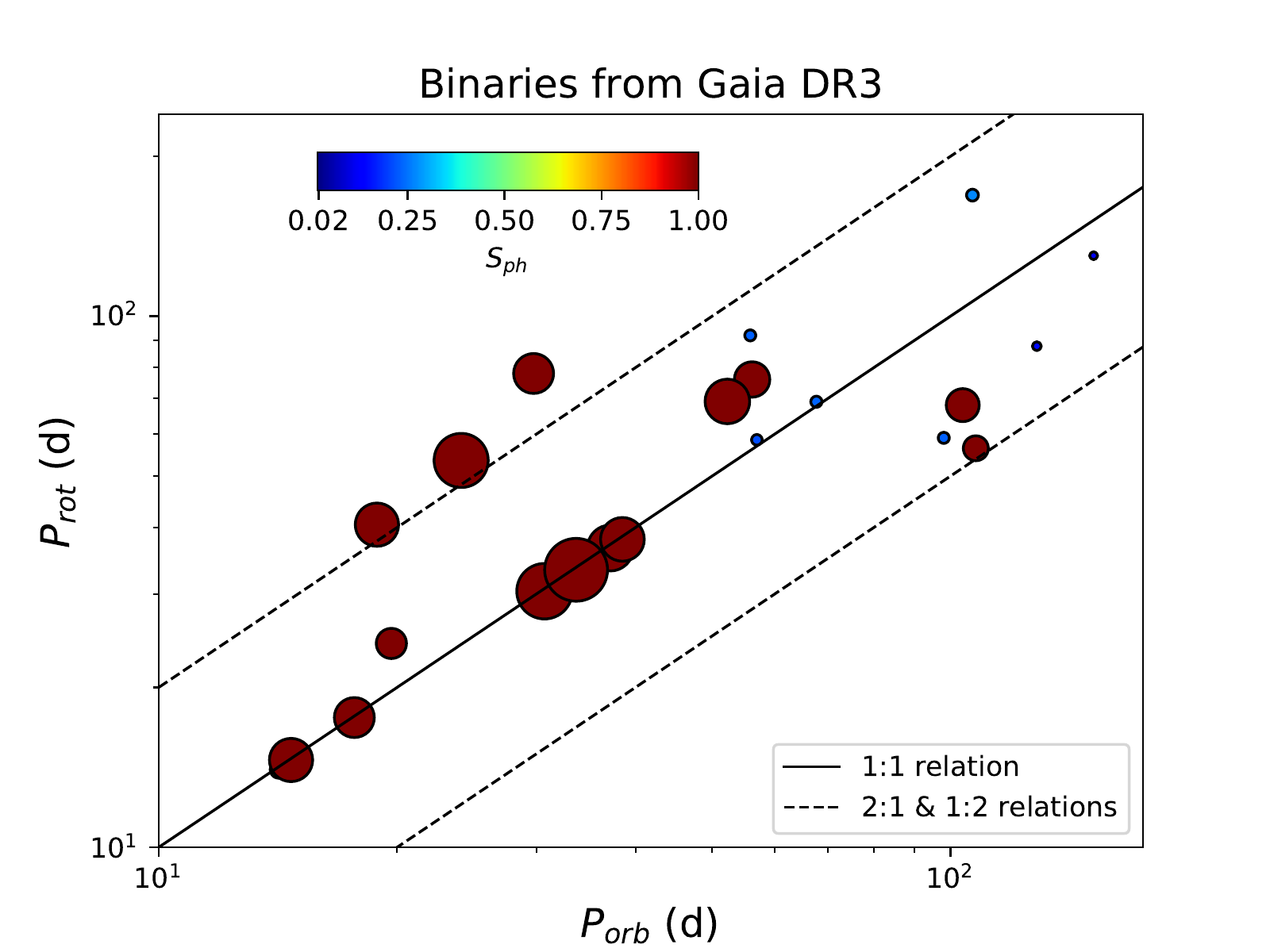}
\caption{\textit{Left:} Rotation period as a function of the orbital period for the close binaries listed by \textit{Gaia} DR3. The symbols’ sizes and color indicate the value of $S\ind{CaII}$. The continuous lines represent the 1:1 relation while the dashed lines represent the 2:1 and 1:2 relations. \textit{Right:} corresponding figure for $S\ind{ph}$ as measured by \citet{Gaulme_2014, Gaulme_2020} and \citet{Benbakoura_2021}.}
\label{fig-Prot-Porb}
\end{figure*}

\section{Close binaries versus single red giants}
\label{Sect_Sindex_binary_single}

\subsection{Close binarity causes enhanced magnetic fields}
If we exclude the stars where no spot modulation was reported, the chromospheric emission and photometric modulation are correlated: a large spot modulation means a large $S\ind{CaII}$ (Fig. \ref{fig-Sph}). We now investigate the impact of close binarity on the chromospheric activity to figure out whether there is a significant difference between single fast-rotating red giants and red giants in close binaries.
Figure \ref{fig-SCa-Prot} (left panel) displays the chromospheric-emission index $S\ind{CaII}$ as a function of the rotational period. As what was found with $S\ind{ph}$ \citep[Fig. \ref{fig-SCa-Prot}, right panel, and][]{Gaulme_2020}, the non-oscillating stars that belong to close binary systems have significantly larger $S\ind{CaII}$ ($\geq 0.25$) than the inactive oscillating giants. They also have values that are significantly larger than the active red giants that are either single or that belong to systems with relatively long orbits. 

We observe a difference between the $S\ind{CaII}$ and $S\ind{ph}$ diagrams (Fig.~\ref{fig-SCa-Prot}): the low activity red giants with partially suppressed oscillations have an $S\ind{ph}$ clearly above the cohort of inactive red giants, while this is not the case for $S\ind{CaII}$. Indeed, $S\ind{CaII}$ is only slightly larger for the active oscillating giants (median value of 0.17) than for the inactive giants (median value of 0.13). This can be seen from the gray shaded areas that indicate the distribution of $S\ind{CaII}$ and $S\ind{ph}$ for the inactive red giants in Fig. \ref{fig-SCa-Prot}. That being said, the main message from Fig. \ref{fig-SCa-Prot} is that $S\ind{CaII}$ is significantly larger in average for stars that belong to close binaries than for single stars and for stars in binaries with long orbital periods, meaning that close binarity leads to larger magnetic fields. The few cases that do not follow this trend are listed in Appendix \ref{App_peculiar_cases}.

Finally, a correlation between chromospheric emission and rotational period has long been reported, by \citet[][]{Noyes_1984} in general, and \citet{Auriere_2015} for red giants. We also observe that $S\ind{CaII}$ decreases as a function of rotational period, but it is mainly due to the stars in close binaries, and is unclear once we remove them. This lack of clear trend probably originates from the small range of spectral types covered in our study, which is limited to K0, G5 and G8 giants. For example, the sample of \citet{Auriere_2015} considers giants with periods from 6 to 600 days, and radii from 5 to 66 $R_\odot$.

\subsection{The key role of spin-orbit resonance}
From a set of 35 red giants in eclipsing binaries, \citet{Benbakoura_2021} reported that the large magnetic activity and its concomitant oscillation suppression appears in systems whose orbits are either in a state of spin-orbit resonance, or of total tidal synchronization.
We verify whether large values of $S\ind{CaII}$ are connected to spin-orbit resonance. 
\citet{Gaulme_2020} report binary status from radial velocity measurements but do not provide orbital periods because of insufficient amount of data. We looked for information about binarity in the last \textit{Gaia} data release \citep[DR3,][]{Gaia_2022_binaries}. Among the 3130 giants for which we have a measured $S\ind{CaII}$, \textit{Gaia} identifies 161 binaries (see also Appendix~\ref{App_Sindex_Porb}), 21 of which are already classified as spectroscopic binaries by \citet{Gaulme_2020}.

Figure \ref{fig-Prot-Porb} (left panel)  displays the rotational versus orbital periods of the close binary systems with $P\ind{orb} \leq 200$ days for which both periods are measured, where the sizes and colors of the symbols indicate the values of $S\ind{CaII}$. We added the active red giants in eclipsing binaries studied by \citet{Gaulme_2014} and \citet{Benbakoura_2021} to our sample. In Fig. \ref{fig-Prot-Porb} (right panel), we represent the corresponding plot for $S\ind{ph}$ as measured by \citet{Gaulme_2014, Gaulme_2020} and \citet{Benbakoura_2021}. Systems with no oscillations that are tidally locked ($1:1$ resonance) or in spin-orbit resonance display the largest values of $S\ind{CaII}$ (median values of 0.36 and 0.46, respectively), compared to the systems that do not have any special tidal configuration (median value of 0.22).
This result is compatible with the trend obtained with $S\ind{ph}$ and confirms that close binarity is not enough to cause large magnetic fields: spin orbit resonance in general, and synchronization and circularization in particular, is the driving factor of abnormally large magnetic fields in red giant stars.

\section{Beyond tidal interaction: looking for signatures of mass transfer and stellar merging}\label{Sect_mass-transfer}

Based on asteroseismic measurements, \cite{Deheuvels_2022} identified intermediate-mass RGB stars with a degenerate helium core, which is unexpected. They interpreted these stars as being initially low-mass red giants with a degenerate core that underwent mass transfer, becoming intermediate-mass giants with a degenerate core. Similarly, \cite{Rui_2021} identified higher-mass RGB stars with a degenerate core, which they propose to be merger remnant candidates, with merging occurring after the formation of a degenerate core on the RGB.

To investigate the impact of such mass gain on the chromospheric activity, we measured the S-index of 34 stars analyzed by \cite{Deheuvels_2022}, including 21 post-mass transfer candidates and 13 control stars, and 17 stars analyzed by \cite{Rui_2021}. The chromospheric emission of the 21 post-mass transfer candidates is not significantly larger than for the control sample of 13 red giants (median $S\ind{CaII}$ of 0.145 versus 0.125). As regards the merger remnant candidates, their chromospheric emission is similar with a median $S\ind{CaII}$ level of 0.134.

We then verified this apparent lack of enhanced magnetic activity from the \textit{Kepler} lightcurves. For the stars that were not part of the \citet{Gaulme_2020} catalog, we downloaded the \textit{Kepler} lightcurves from the Mikulski Archive for Space Telescopes (MAST) and processed them with the same pipeline. We found no significant spot modulation ($S\ind{ph}\leq0.1\,\%$) for the stars of these two samples. Therefore, we conclude that both samples are magnetically inactive. In other words, these stars have not experienced any significant angular momentum enhancement able to trigger a dynamo mechanism. However, the non-detection of magnetic activity may reflect a selection bias as stated by \cite{Rui_2021},  because the two studies focus on red giants with high signal-to-noise ratio (SNR) oscillations. A stronger magnetic activity would result in partially or totally suppressed oscillations, for which their analysis based on interpreting the mixed modes would be compromised.

\section{Conclusions}\label{Sect_conclusion}

We measured the chromospheric activity of 3130 low-red giant branch and red clump stars observed by \textit{Kepler} and analyzed by \cite{Gaulme_2020}, using data from the LAMOST survey (DR7). For stars where spots were detected from the \textit{Kepler} lightcurves, we verified that there is a correlation between the chromospheric emission index ($S$-index) $S\ind{CaII}$ and the photometric index $S\ind{ph}$. This indicates that the intensity of spot modulation is proportional in average to the strength of surface magnetic fields, even though stellar inclination can bias this trend. We also noticed that $S\ind{CaII}$ is correlated with $\log g$ for quiet red giants, that is, where no spots are detected. 

Regarding the main objective of this work, we conclude that red giants in close binaries that are in a configuration of spin-orbit resonance develop larger magnetic fields than single red giants with similar rotation rates. In other words, the large magnetic field of red giants in close binary systems is not only due to the faster rotation rate induced by tidal interactions, as it has been generally admitted for RS CVn stars \citep[e.g.,][]{Shore_1994}. Somehow, our work resuscitates an old speculation about a special binary-induced dynamo activity \citep[e.g.,][]{Hall_1976}, which was discarded a few years later by \citet{Morgan_Eggleton_1979} who interpreted it as the result of a selection bias when less than 30 RS CVn stars were known.

In a broader context, we looked for signatures of stellar mergers or post-mass transfer reported by \citet{Rui_2021} and \citet{Deheuvels_2022}, but we did not obtain any significant outcome from $S\ind{CaII}$ and $S\ind{ph}$ measurements. We also looked for a correlation between chromospheric emission and rotational period, as was observed by e.g., \citet{Auriere_2015}. We do not confirm their results, likely because of the much smaller range of spectral types covered by our study.

\begin{acknowledgements}
C.G. was supported by Max Planck Society (Max Planck Gesellschaft) grant “Preparations for PLATO Science” M.FE.A.Aero 0011. P.G. was supported by the German space agency (Deutsches Zentrum für Luft- und Raumfahrt) under PLATO data grant 50OO1501.
J.Y. acknowledge support from ERC Synergy Grant WHOLE SUN 810218. The authors thank J. Gomes da Silva for providing the HARPS spectra analyzed in this study, as well as J. Zhang for the fruitful discussion. Guoshoujing Telescope (the Large Sky Area Multi-Object Fiber Spectroscopic Telescope LAMOST) is a National Major Scientific Project built by the Chinese Academy of Sciences. Funding for the project has been provided by the National Development and Reform Commission. LAMOST is operated and managed by the National Astronomical Observatories, Chinese Academy of Sciences. P.G. and C.G. hope that Pr. B. Mosser will enjoy reading the first work led by two of his former PhD students.
\end{acknowledgements}

\bibliographystyle{./bibtex/aa} 
\bibliography{./bibtex/biblio}


 \begin{appendix}
 
\section{Median values of $S\ind{CaII}$ for different categories of red giants}\label{App_median_S_index}

In Table \ref{table:2}, we report the median S-index measured for each category of red giants listed by \cite{Gaulme_2020}. We note from Fig. \ref{fig-Sph} that the clearer the detection of the photometric modulation done by \citet{Gaulme_2020}, the larger the S-index in average. We consistently obtain increasing median S-indices for the non-active, the low-SNR active and the clearly active sample, respectively.

\begin{table}
\centering
\caption{Median $S$-index measured for different categories of red giants.}
\begin{tabular}{c c}
\hline
Group & Median $S\ind{CaII}$ \\ 
\hline\hline
\multicolumn{2}{c}{Red giants from \cite{Gaulme_2020}}\\
\hline
Inactive & 0.133 \\
Low SNR spots & 0.161\\ 
Clear spots & 0.175\\
\hline
\hline
\multicolumn{2}{c}{\textit{Gaia} DR3 binaries}\\
\hline
All binaries &  0.154 \\
Wide binaries & 0.149 \\
Close binaries & 0.165 \\
Non-osc. \& spin-orbit res. & 0.410 \\
\hline
\hline
\multicolumn{2}{c}{Post mass-transfer candidates}\\
\hline
Control sample &  0.125\\
Post-mass transfer candidates &0.145 \\
Merger remnant candidates & 0.134\\
\hline
\end{tabular}
\tablefoot{Block 1: red giants from \cite{Gaulme_2020}. Inactive stars present no evidence for rotational modulation in the light curve, while stars with low-SNR spots and clear spots exhibit rotational modulation and have a measurement of their rotation period. Block 2: red giants from \cite{Gaulme_2020} that are identified as binaries by \textit{Gaia} DR3. Wide binaries have an orbital period $P\ind{orb} > 150$ days while close binaries have $P\ind{orb} \leq 150$ days. Non-oscillating stars are in a close binary system that is tidally locked or in spin-orbit resonance, with clear rotational modulation and a measured rotation period. Block 3: post-mass transfer red giant candidates of \cite{Deheuvels_2022} and \cite{Rui_2021}. The control sample of \cite{Deheuvels_2022} is composed of RGB stars with no evidence for a mass gain, i.e. low-mass red giants with a degenerate core and intermediate-mass red giants with a non-degenerate core.}
\label{table:2}
\end{table}

\section{Peculiar cases}
\label{App_peculiar_cases}
From Figs. \ref{fig-SCa-Prot} and \ref{fig-Prot-Porb} it is clear that red giants that belong to close binary systems and are in a configuration of spin-orbit resonance lead to enhanced magnetic fields with respect to single stars with similar rotational periods, and to binary systems that do not have any special tidal configuration. However, a few cases stand apart. 

We note that five oscillating red giants with $P\ind{rot} < 80$ days present $S\ind{CaII} > 0.25$ (Fig. \ref{fig-SCa-Prot}). Among these, two are RC stars showing no evidence for a binary signal in the power spectrum of their light curve (KIC 5372141, 6791309). They might have undergone special events during the previous RGB phase resulting in an enhanced activity, such as planet engulfment \citep{Privitera_2016a, Privitera_2016b, Tayar_2022}, mass transfer \citep{Deheuvels_2022} or a stellar merger \citep{Rui_2021}. Investigating the origin of the enhanced magnetic activity for these two RC stars is beyond the scope of this work. The evolutionary stage of the three remaining red giants with $P\ind{rot} < 80$ days and $S\ind{CaII} > 0.25$ is unidentified (KIC 5112741, 5471548,  10264774). They exhibit partially suppressed oscillations, which did not allow \cite{Gaulme_2020} to identify whether these are RGB or RC stars. \cite{Gaulme_2020} found that 15\,\% red giants with partially suppressed oscillations belong to binaries. It may be the case of these three stars, and more data are required to investigate whether they belong to a binary system.

\section{Orbital periods of active red giants}\label{App_Sindex_Porb}

\begin{figure}
\centering
\includegraphics[width=9.1cm]{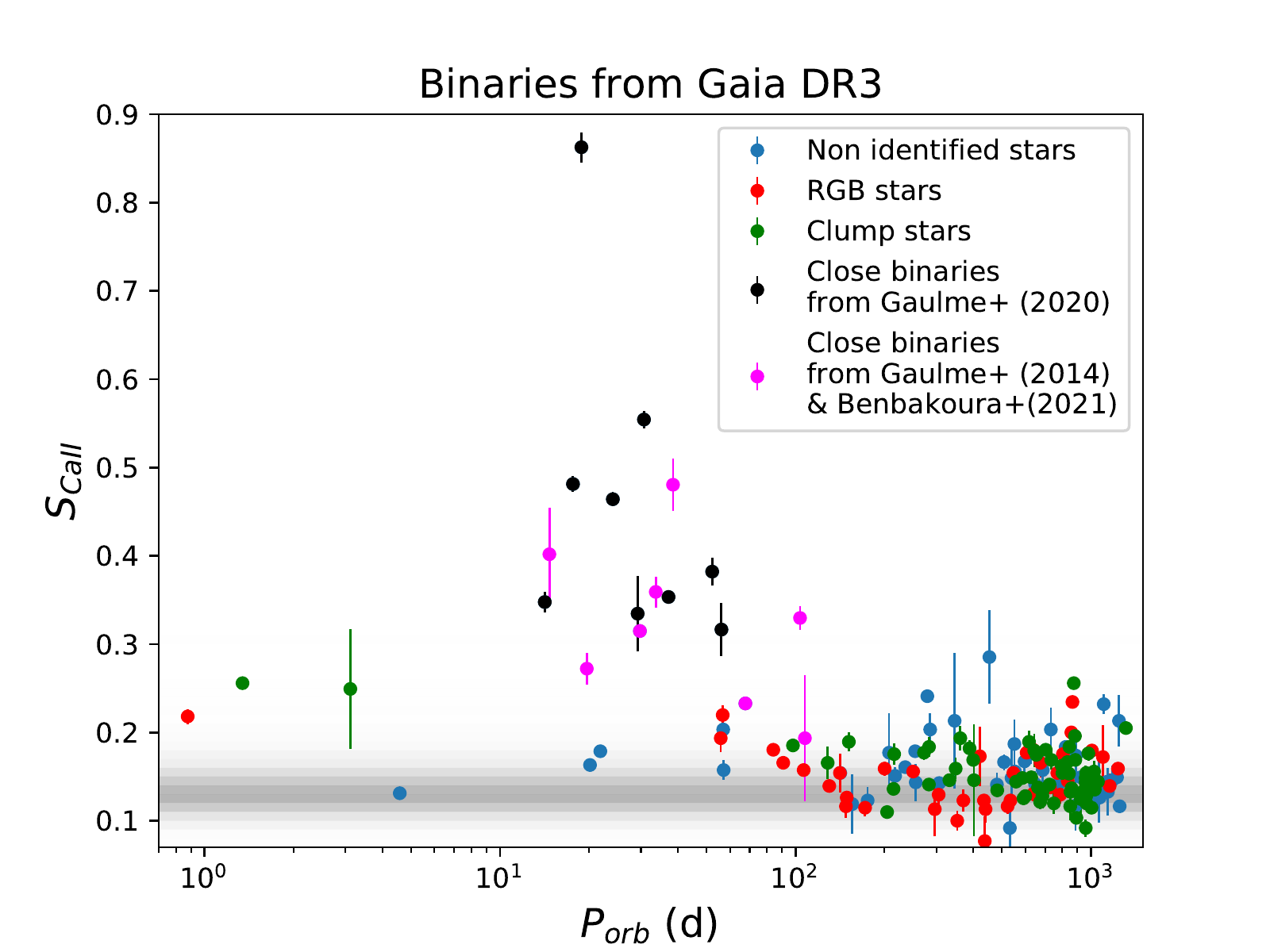}
\includegraphics[width=9.1cm]{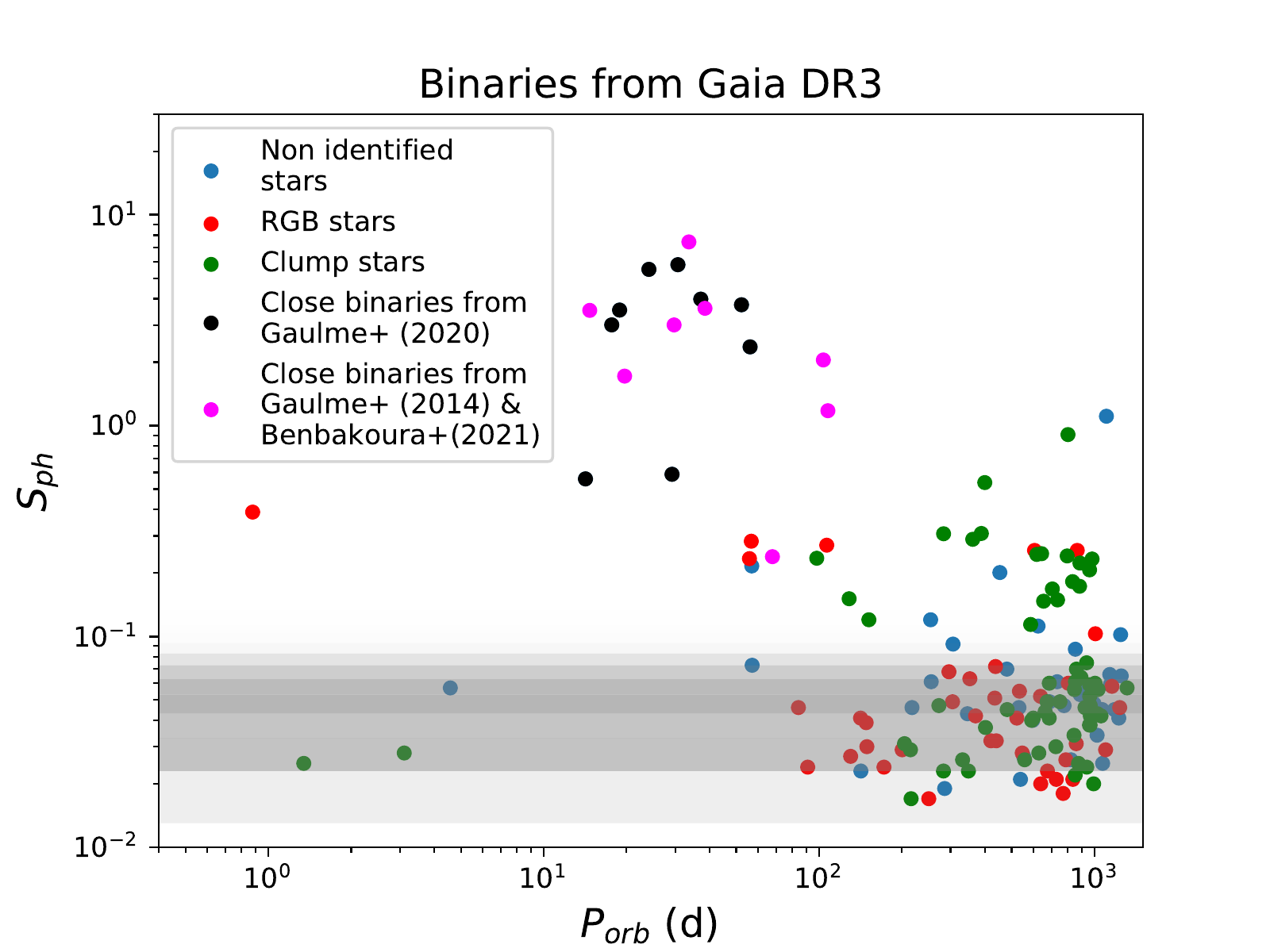}
\caption{Magnetic activity of red giants in binaries from \cite{Gaulme_2020} identified by \textit{Gaia} DR3 as a function of the orbital period. RGB stars are represented in red, RC stars in green, red giants with an unidentified evolutionary stage in blue, non-oscillating red giants from \citet{Gaulme_2020} in black, and red giants from \citet{Gaulme_2014, Benbakoura_2021} in magenta.
Top panel: S-index $S\ind{CaII}$. Bottom panel: Photometric index $S\ind{ph}$ as measured by \citet{Gaulme_2020}.}
\label{fig-SCa-Porb}
\end{figure}

To evaluate whether some red giants belong to binary systems that were missed by \citet{Gaulme_2020} we looked for possible binaries from the \textit{Gaia} DR3 catalog. We found 161 stars that seem to belong to binary systems, with orbital periods up to about 1000 days. Besides, to study the importance of spin-orbit resonance, we added the red giants in eclipsing binaries that were studied by \citet{Gaulme_2014} and \citet{Benbakoura_2021}. Fig. \ref{fig-SCa-Porb} displays both $S\ind{CaII}$ and $S\ind{ph}$ as a function of their orbital periods. As expected, we see that both $S\ind{CaII}$ and $S\ind{ph}$ depend on the orbital period. However, this figure also reminds us to not take everything for granted when looking at data from large catalogs, \textit{Gaia} and \textit{Kepler} in our case. For example, the targets KICs 4543371, 4761037, 5701850, and 7670875  appear to be binary systems with periods shorter than 5 days and a low surface magnetism ($S\ind{CaII} < 0.25$ and $S\ind{ph} < 0.4\,\%$). Given that their asteroseismic radii are of about 8.4, 8.3, 10.6, and 9.9 $R_\odot$, respectively, these ``systems'' are likely the result of photometrical blending between short-orbit binaries and red giant stars on the \textit{Kepler} detector, or triple systems where the red giant component is on a longer orbit.

\end{appendix}

\end{document}